\documentclass[fleqn,10pt]{wlscirep}
\title{A new bound on polymer quantization via an opto-mechanical setup}

\author[1]{Mohsen Khodadi}
\author[1,*]{Kourosh Nozari}
\author[2]{Sanjib Dey}
\author[3]{Anha Bhat}
\author[4,5]{Mir Faizal}
\affil[1]{Department of Physics, Faculty of Basic Sciences,
University of Mazandaran, P. O. Box 47416-95447, Babolsar, Iran}
\affil[2]{Department of Physics, Indian Institute of Science Education and Research Mohali, Sector 81, SAS Nagar, Manauli 140306, India}
\affil[3]{Department of Metallurgical and Materials Engineering, National Institute of \protect Technology, Srinagar 190006, India}
\affil[4]{Irving K. Barber School of Arts and Sciences, University of British Columbia-Okanagan,\protect 3333 University Way, Kelowna, British Columbia V1V 1V7, Canada}
\affil[5]{Department of Physics and Astronomy, University of Lethbridge, Lethbridge, Alberta T1K 3M4, Canada}
\affil[*]{Corresponding author email: knozari@umz.ac.ir}

\begin{abstract}
The existence of a minimal measurable length as a characteristic length in the Planck scale is one of the main features of quantum gravity and has been widely explored in the context. Various different deformations of spacetime have been employed successfully for the purpose. However, polymer quantization approach is a relatively new and dynamic field towards the quantum gravity phenomenology, which emerges from the symmetric sector of the loop quantum gravity. In this article, we extend the standard ideas of polymer quantization to find a new and tighter bound on the polymer deformation parameter. Our protocol relies on an opto-mechanical experimental setup that was originally proposed in Ref.\cite{ref:Igor} to explore some interesting phenomena by embedding the minimal length into the standard canonical commutation relation. We extend this scheme to probe the \emph{polymer length} deformed canonical commutation relation of the center of mass mode of a mechanical oscillator with a mass around the Planck scale. The method utilizes the novelty of exchanging the relevant mechanical  information with a high intensity optical pulse inside an optical cavity. We also demonstrate that our proposal is within the reach of the current technologies and, thus, it could uncover a decent realization of quantum gravitational phenomena thorough a simple table-top experiment.
\end{abstract}
\begin{document}

\flushbottom
\maketitle
%
%


\section*{Introduction}

After more than 70 years of focusing on the theoretical and mathematical aspects of the theory of quantum gravity (QG),
in recent decades we have encountered some serious proposals on the project of the QG phenomenology \cite{ref:Stachel}.
Indeed, the first phenomenological proposition for QG from the experimental point of view has been set up very recently
\cite{ref:Amelino0}. This has given rise to the  possibility of bringing other areas like astrophysics, particle physics,
cosmology, together in a same footing along with some feasibilities for experiment. Thereby, guided by the empirical facts
into the context of different approaches of QG, the novel concepts are growing rapidly day by day \cite{ref:Amelino1}.
Today, we are dealing with deliberately genuine schemes of QG phenomenology in different contexts like string theory
\cite{ref:st1,ref:st2}, loop quantum gravity (LQG) \cite{ref:lqg1,ref:lqg2}, doubly special relativity \cite{ref:dsr1,ref:dsr2},
etc., which are not only successful from the theoretical aspects but also they made a significant impact on relevant experiments \cite{ref:Sabin}.

Several approaches of QG including those mentioned above support the fact that in the QG regime the space-time has a discrete structure
and the Planck length is the smallest measurable and invariant length with respect to all inertial observers in the nature,
which plays the role of a regulator for quantum field theories. In ordinary Schr\"odinger representation of quantum mechanics,
the spectra of the position and momentum operators being continuous, such natural cut-off can not be implemented by using
the standard quantum mechanical theory. However, one can present an effective model
of QG to introduce the Planck length through the modification of the Heisenberg' uncertainty principle.
More specifically, given that gravity is not an ordinary force, but a property of space-time, we can deal
with an effective framework of QG by considering a fundamental and minimal characteristic length for the geometry
of space-time that is probed by a moving quantum particle. Within such context the standard Heisenberg uncertainty
principle and relativistic dispersion relation are replaced by the \textit{generalized uncertainty principle} (GUP)
\cite{ref:Maggiro, ref:Kempf1, ref:Kempf2, ref:Ali,ref:Gomes,ref:Bagchi,ref:Quesne,ref:Pedram} and
the \textit{modified dispersion relation} (MDR) \cite{ref:Amelino,ref:Majhi}, respectively, where the geometry
of momentum space is not trivial anymore and sometimes the Lorentz symmetry may also be broken.

Polymer quantum mechanics is one of the most recent phenomenological approaches to the problem of QG, which is suggested in the symmetric sector
of LQG \cite{ref:Ashtekar}. The original version of polymer quantum mechanics is a quantum mechanics on a lattice such that the
lattice length plays the role of a minimal length \cite{ref:Corichi1}. In standard quantum mechanical theory, the position and momentum
operators are unbounded operators, however, in order to insert a natural cut-off in the quantum gravity regime, one requires to deal with
the bounded observables. Polymer quantum mechanics is an excellent framework to achieve this
\cite{ref:Kunstatter,ref:Laddha,ref:Hossain,ref:Prieto}. It has also been demonstrated that polymer quantization and GUP
have the same physical consiquences \cite{ref:Majumder}.
Within this scheme, one implements the Weyl algebra \cite{ref:Ashtekar,ref:Corichi1} by preserving the standard canonical form of the commutation relations such that the momentum operator becomes ill-defined. However, it can be regularized later by introducing an exponential shift operator so that the momentum operator becomes well-defined \cite{ref:Ashtekar}. As a consequence, one obtains a modified version of the momentum operator and, hence, the Hamiltonian, which supports the existence of a minimal length. Meanwhile, an alternative scheme for this also exists. One can implement a noncanonical representation of the Heisenberg algebra such that the Hamiltonian operator remains in its standard functional form but the commutation relation is modified as \cite{ref:GUP-Polymer1, ref:GUP-Polymer2}
\begin{equation}\label{e2-00}
[x,p]_{\mu_0}=i\hbar \sqrt{1-\mu_0^2\left(\frac{p}{M_\text{P}c}\right)^2}.
\end{equation}
Here $M_\text{P}\sim 10^{19}$ GeV is the Planck mass and $\mu_0$ labels a numerical coefficient which determines the precise value of the polymer length $\mu_0 l_P$. It is expected that $\mu_0$ should be of the order of unity $\mu_0={\mathcal O}(1)$, but its precise value can only be determined by experiments. In this paper, we are interested to constrain this quantity through an opto-mechanical (OM) experimental setup. It should be noted that there exist several different phenomenological aspects to test the short distance effects of QG \cite{ref:Ali,ref:Das,ref:Pedram,ref:Nozari}, however, a more recent approach based on an OM experimental scheme serves a more interesting way to the problem \cite{ref:Igor}. The biggest advantage of the method discussed in the latter scenario is that one does not require an expansive high energy scattering experiment to probe the Planck length accuracy in position measurement, which is not possible anyway since the highest energy possible to reach by the resources available to us is about 15 orders of magnitude away from the Planck energy \cite{ref:Amelino2}. The system \cite{ref:Igor}, rather, based on a simple table-top experiment which can be utilized cleverly to serve the purpose. Being inspired by the optical layout designed in \cite{ref:Igor}, we setup a similar scheme to test the polymer deformation, which emerges from the symmetric sector of LQG, while in \cite{ref:Igor} the authors explored the existence of minimal length in QG instead. One of the interesting results in our article is that we notice that the polymer deformation parameter $\mu_0$ can be measured with a remarkable sensitivity under this scheme, which we discuss in the following section. Moreover, it should be emphasized that following the GUP model considered in \cite{ref:Igor}, we apply the polymer modified commutation relation (\ref{e2-00}) to the \emph{center of mass} rather than to each single particle individually. Otherwise, the underlying OM scheme would not be able to put a stringent bound on $\mu_0$.

\section{Opto-mechanical scheme for polymer-modified commutation relation}\label{sec2}
To start with, let us first represent the position and momentum operators $x,p$ corresponding to our system (\ref{e2-00}) as dimensionless observables $X_m=x\sqrt{m\omega_m/\hbar},P_m=p/\sqrt{\hbar m\omega_m}$, which are familiar as quadratures in quantum optics. Consequently, the polymer modified commutation relation (\ref{e2-00}) obtains the following form
\begin{equation}\label{e3-00}
[X_m,P_m]_{\mu_0}=i\hbar \sqrt{1-\mu^2P_m^2},
\end{equation}
with $\mu=\mu_0\sqrt{\hbar m\omega_m}/M_\text{P}c$. The OM scheme is designed in a way that the mechanical oscillator of mass $m$ and angular frequency $\omega_m$ (\ref{e3-00}) interacts with an optical pulse in a very efficient way inside an optical cavity. In order to understand this interaction in reality, one requires a unitary displacement operator \cite{ref:Vanner} $U_m=e^{in_L\lambda X_m}$, which displaces the quadrature $X_m$ of a mechanical oscillator in phase space induced by the optical field of interaction length $\lambda$. Here $n_L$ represents the photon number operator. A sequence of four such radiation pressure interactions causes the mechanical state to be displaced around a loop in phase space, which effectively forms an optical cavity (resonator) yielding the total interaction operator
\begin{equation}\label{e3-1}
\xi = e^{ in_L\lambda P_m} \,  e^{- in_L\lambda X_m}
\, e^{ -in_L\lambda P_m} \, e^{ in_L\lambda X_m} \, .
\end{equation}
After a complete sequence, apparently it seems that none of the two systems would be affected, since the operations are likely to neutralize the effects of each other classically. However, interestingly, it was argued in \cite{ref:Amart2016} that both the classical and quantum mechanical cases contribute to the effect apart from some differences in values. So, in principle one can consider any of the effects for the purpose. However, since our work is based on the quantum mechanical scenario, it is more natural that we consider the quantum effects only. Consequently, the two subsequent displacements in phase space governed by (\ref{e3-1}) cause an additional phase to the state under consideration, especially when the commutation relation between $X_m$ and $P_m$  is modified as (\ref{e3-00}). This additional phase shift in the oscillator will create a change in the optical field correspondingly, which is what we need to measure. Thus, a finite shift in the optical field will confirm the existence of the polymer modified commutation relation. In order to obtain the desired change in the optical field, let us analyze the mean of the optical field operator
\begin{equation}\label{e3-3Polyapp}
\left\langle a_L \right\rangle =\langle\alpha\vert\xi^\dagger a\xi\vert\alpha\rangle =\left\langle a_L \right\rangle_{\text{QM}} e^{- i \Theta},
\end{equation}
where $\vert\alpha\rangle$ is the coherent state of the input optical field with mean photon number $N_p$, $a_L$ being the annihilation operator and $\Theta$ describes the additional optical phase emerging due to the polymer modified commutation relation. Meanwhile, the average of the standard quantum mechanical field operator can be computed as \cite{ref:Bosso}
\begin{equation}
\left\langle a_L \right\rangle_{qm}= e^{ -i \lambda^2 -N_p
\left(1- e^{-2i \lambda^2} \right)}.
\end{equation}
Now, the only thing that remains is to compute $\langle\alpha\vert\xi^\dagger a\xi\vert\alpha\rangle$ as described in (\ref{e3-3Polyapp}). For this, we first simplify the interaction operator (\ref{e3-1}). Using the standard relation $e^{aA}Be^{-aA}=\sum_{k=0}^\infty \frac{i^k a^k}{k!} c_k$, with $i c_k = [A,c_{k-1}]$ and $c_0= B$, one can re-express the total interaction operator (\ref{e3-1}) as follows
\begin{equation}\label{e3-2}
\xi = e^{- in_L\lambda\sum_{k} \frac{c_k\left(
\lambda n_L \right)^k }{k!}}.
\end{equation}
This relation clearly depends on the commutation relation related to the mechanical oscillator, but not with that of the optical field. It is easy to demonstrate that within the context of the standard quantum mechanics one achieves $\xi = e^{-i \lambda^2 n_L^2 }$ \cite{ref:Igor}. This means that in the absence of QG, the optical field is affected only by a self-Kerr-nonlinearity term as $n_L^2$ operation and the mechanical state stays
unchanged. However, it is clear that $c_k$ with $k>1$ are nonzero in polymer setup, while all of them are vanishing in the standard quantum mechanics. Considering $A=X_m$ and $B=c_0=P_m$, it is easy to show that all the coefficients $c_k$ with $k>1$ are nonzero and can be represented by the recursive relation as
\begin{equation}\label{ck}
c_k=-\mu\,c_{k-2},\qquad k=2,3,....,
\end{equation}
with
\begin{equation}\label{c0-c1}
c_0=P_m,\qquad c_1=\sqrt{1-\mu^2P_m^2} .
\end{equation}
This implies that all the coefficients $c_k$ are determined only by two independent entities $c_0$ and $c_1$. Therefore, we can simply write the displacement operator (\ref{e3-2}) in terms of these two independent coefficients as
\begin{equation}\label{e3-2Poly}
\xi = e^{- i \lambda n_L \left(-\mu \frac{\lambda^2n_L^2}{2!}
\right) c_0 - i \lambda^2 n_L^2 \left(1-\mu \frac{\lambda^2 n_L^2
}{3!}\right) c_1},
\end{equation}
which can be rewritten in a simpler form as
\begin{equation}\label{e3-2Polyapp}
\xi = e^{- i \lambda^2 n_{L}^2 c_1}
e^{i \mu \big(\frac{\lambda^3 n_L^3}{2} P_m+
\frac{\lambda^4 n_L^4}{6} c_1\big)}.
\end{equation}
Note that, we have not considered the contribution of $c_k$ for $k\geq4$, since these terms have negligible effects for the order of ${\mathcal O}(\mu^2)$ and higher. As we discussed earlier, in the absence of the polymer deformation parameter, i.e. for $\mu=0$, the above equation reduces to the relation corresponding to the standard quantum mechanics. However, for the case considered here, we notice a deviation from that, which is given by the second exponential. Consequently, from (\ref{e3-3Polyapp}) we compute the deviation in the average value of the field operator with $c_1\approx 1$ and follow the similar steps as in \cite{ref:Igor, ref:Bosso} to obtain the deviation of the optical phase as
\begin{equation}\label{e3-5Polyapp}
\Theta(\mu)  \simeq \frac{2}{3} \mu N_p^3 \lambda^4  \, e^{-i n \lambda^2},
\end{equation}
\begin{figure}
\includegraphics[scale=0.9]{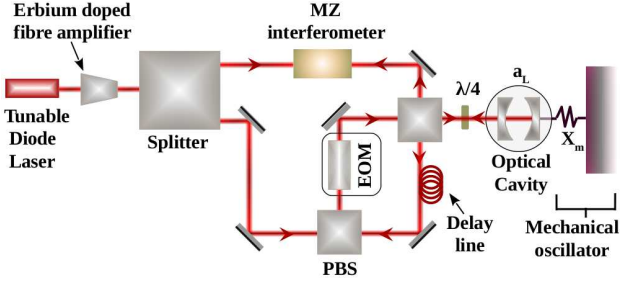}
\caption{Experimental setup inspired by \cite{ref:Igor} to probe the induced effects of \emph{polymer modified commutation relation} in a macroscopic mechanical resonator.}
\label{Fig1}
\end{figure}
\section{Schematic for experimental setup}\label{sec3}
Let us now discuss a realistic experimental scenario that can measure the additional phase (\ref{e3-5Polyapp}) emerging form the polymer modified commutation relation. The OM scheme that we shall explore here have been utilized for different purposes \cite{ref:Igor,ref:Kippenberg, ref:Connell,ref:Teufel,ref:Grossardt,ref:Gan,Dey_NPB}. In our case, the device couples the polymer modified mechanical oscillator with an optical pulse via radiation pressure inside a high-finesse optical cavity as depicted in Fig. \ref{Fig1}. More specifically, the input light is created from a tunable diode laser and, subsequently, amplified by an erbium doped fibre amplifier. Then, it is divided into two parts, one part is kept as a reference beam for future use. While the other is passed through a polarized beam splitter (PBS) followed by an electro-optic modulator (EOM) to allow for an interaction with a mechanical oscillator having position $X_m$ inside a high-finesse optical cavity with cavity field $a_L$. The light is then retro-reflected and enters into the delay line with a vertical polarization. After interacting with the PBS again, its polarization turns into the horizontal direction so that it can interact with the mechanical oscillator again. The same process is repeated four times in total so that the canonical commutator is mapped onto the optical field as described in (\ref{e3-1}). Finally, the EOM is operated in such a way that it does not rotate the polarization and the light is taken out of the system. The output beam, thus, contains all the information of the OM interaction and, therefore, it must possess an additional phase $\Theta$ as described before in (\ref{e3-5Polyapp}). This will be evident when it is tested with respect to the reference beam by any standard interferometer like Mach-Zehnder (MZ) interferometer.

The EOM that we use here has to be very efficient. We design it in a way that it uses a particular type of electro-optic effect (Pockels effect) to modulate the phase of the incident beam of light. We use a refractive modulator (whose refractive index changes due to the application of electric signal that causes the modulation) that produces an output beam of light whose phase is modulated with the electric signal applied to the Beta Barium borate (BBO) electro-optic crystal. The modulation can be controlled by the source of electric signal according to our requirement.

\section{A new tight constraint on the polymer deformation parameter $\mu_0$}\label{sec4}
Let us now realize how the above scheme can be implemented in reality with the available resources in a standard optics lab. The above OM interaction can be described by the inter-cavity Hamiltonian $H = \hbar \omega_m n_m - \hbar g_0 n_L X_m$, where $n_m$ is the mechanical number operator and $g_0 = \omega_c \sqrt{\hbar}/(L\sqrt{m\omega_m})$ denotes OM coupling rate with the mean cavity frequency $\omega_c$ and mean cavity length $L$ \cite{ref:Law1995}. By assuming the optical pulses to be short enough, one can approximate the intra-cavity dynamics via the unitary operation $U = e^{i \lambda n_L X_m} $ \cite{ref:Vanner2010}. Here $\lambda \simeq g_0/\kappa=4 \mathcal{F}\sqrt{\hbar}/(\lambda_L\sqrt{m\omega_m})$ refers to the effective interaction strength in which $\kappa$ and $\lambda_L$ are the amplitude decay rate and the wavelength of optical pluses, respectively, with the cavity finesse $\mathcal{F}$ \cite{ref:Igor}. In this framework, the OM phase (\ref{e3-5Polyapp}) is modified as
\begin{equation}\label{e3-6Polyapp}
\Theta= \frac{512\mu_0 \hbar^{5/2}{\mathcal F}^{4} N_{p}^{3}}{3M_{\text{P}}c \lambda_L^4(m \omega_m)^{3/2}}.
\end{equation}
In realistic scenarios, we face many different sources of background noise which may limit the ability to detect our measurable quantity in the laboratory. Thus, it is essential to study the signal to noise ratio (SNR) for our system, SNR$=\Theta/\delta\Phi$, where $\delta\Phi$ is the uncertainty in measuring the optical phase shift. For a proper experimental setup, it is required that SNR$>1$. The inaccuracy of the measurement $\delta\Phi$ depends on the quantum noise of the outgoing optical pulse $\sigma_{\text{out}}$. In an ideal experiment with coherent state of light of mean photon number $N_p$, the phase uncertainty turns out to be $\delta\Phi=\sigma_{\text{out}}/\sqrt{N_pN_r}$, where $N_r$ is the number of independent runs of the experiment. Therefore, the precision of the measurements are not limited and can be enhanced further by adjusting the strength of the optical field $N_p$ as well as the number of experimental runs $N_r$, from which one can directly compute the resolution of $\delta\mu_0$. For every measurement, there exists a realistic parameter regime of $\mu_0$. For instance, let us to set the values $\omega_m =2 \pi\times 10^5$~Hz, $m=10^{-11}$~kg and ${\mathcal F}=10^5$ which are the mechanical oscillator frequency, oscillator mass and optical cavity of finesse, respectively, with a wavelength of $\lambda_L=1064$~nm, all of which are within the the range of current experiments \cite{ref:Corbitt2007,ref:Harris2008,ref:Verlot2008,ref:Groblacher2009,Kleckner}. Now by performing a single run of the experiment, i.e. $N_r=1$, as well as by fixing the pulse sequence of mean photon number $N_p=10^8$, within a relevant basic inaccuracy $\delta\Phi$ we reach at $\delta\mu_0\sim10^4$ for the resolution of the dimensionless polymer modified parameter. Interestingly, by decreasing the mechanical oscillator frequency and oscillator mass to $\omega_m =2 \pi\times
10^3$~Hz, $m=10^{-13}$~kg, respectively, while other involving experimental parameters remain unchanged, one finds $\delta\mu_0\sim10$, which is very tighter than the formerly obtained results. Also by increasing the photon number as well as the number of measurement runs to $N_p=10^{10}$ and $N_r=10^2~,10^4,~10^6$, one acquires very impressive resolution of the order of magnitudes $\delta\mu_0\sim 10^2,~10,~1$, respectively. However, it should be noted that the above experimental parameters are strongly affected by disorders such as \emph{noise sources}, \emph{mechanical damping}, etc., which are subject to more technical discussion and, therefore, we refer the readers to the supplementary information provided in \cite{ref:Igor}. Some detailed experimental analysis has also been performed in the spirit of the underlying scheme in Refs. \cite{ref:Martin2014, ref:Bawj2015, ref:Kumar2017}, which provide a deeper understanding of the subject. Although at the first glance it seems to be challenging to achieve some values attributable to the above experimental parameters, however, as we showed with the resources available to us it is not that difficult. In fact, with some fine adjustments in our experimental setup and/or parameters, it may also be possible to reach an impressive resolution where the polymer length deformation may turn out to be of the order of Planck length. However, that requires a lot more sophistication.

\section{Conclusions}\label{sec5}
We have explored a detailed and elegant procedure to understand the effects of the modified commutation relation thorough polymer quantization in laboratory. Our method utilizes an OM setup designed originally in \cite{ref:Igor} that helps us to transfer the information of a polymer modified mechanical oscillator to the high intensity optical pulse in terms of a sequence of OM interaction inside an optical resonator. Consequently, we end up with an optical phase shift that is easily measurable with a very high accuracy through a MZ interferometric system. This makes the whole procedure much easier to collect the information of polymer deformations thorough an elegant optical system already available to us. Moreover, we obtained a new bound on the deformation parameter $\mu_0$, which may lead us towards an advanced understanding of the polymer quantization as well as the problem of quantum gravity.
In what follows, we point out our contribution in comparison to the seminal work \cite{ref:Igor} conducted on the framework of GUP models. While the Ref.\cite{ref:Igor} is in the framework of GUP, our study is based on polymer quantization scheme which is a relatively new and dynamic field towards the quantum gravity phenomenology. Although, our modified commutation relation (\ref{e2-00}), which addresses a maximum value for the momentum as $p<M_{p}/\mu_{0}$ have also been studied in the context of GUP, however, they are not conceptually equivalent in QG ground. In fact, the origin of the fundamental length scale arising from the frameworks of polymer quantization and GUP are different, while GUP setups are suggested phenomenologically, our setup emerges naturally in the non-relativistic limit of the symmetric sector of LQG. Thus, one may argue that the polymer setup is more appealing at least in this respect. Furthermore, our contribution explicitly shows that the polymer length, like the fundamental length scale emerging from the GUP schemes, can be detected with a remarkable resolution as expected from the theory via some fine adjustments on the experimental parameters through the OM scheme proposed in \cite{ref:Igor}.\\

\section*{Acknowledgements}

The authors would like to thank M. A. Gorji for his insightful discussion and comments. S.D. is supported by an INSPIRE Faculty Grant (DST/INSPIRE/04/2016/001391) by the Department of Science and Technology, Government of India. A.B. is supported by MHRD, Government of India and would like to thank Department of Physics and MMED at NIT Srinagar for carrying out her research pursuit.\\

\bibliography{sample}

\begin{thebibliography}{1}
\bibitem{ref:Stachel}
J. Stachel, \emph{Early history of quantum gravity, Black Holes, Gravitational Radiation and the Universe}, edited by B. Iyer and B. Bhawal, Kluwer Academic Publisher: Dordrecht (1999).

\bibitem{ref:Amelino0}
G. Amelino-Camelia, Quantum theory's last challenge, Nature \textbf{408}, 661--664 (2000).

\bibitem{ref:Amelino1}
G. Amelino-Camelia, Quantum-spacetime phenomenology, Living Rev. Rel. \textbf{16} 5, (2013).
\bibitem{ref:st1}
D. J. Gross and P. F. Mende, String theory beyond the Planck scale, Nucl. Phys. B \textbf{303} 407--454 (1988).

\bibitem{ref:st2}
D. Amati, M. Ciafaloni and G. Veneziano, Can spacetime be probed below the string size? Phys. Lett. B \textbf{216}, 41--47 (1989).

\bibitem{ref:lqg1}
C. Rovelli and L. Smolin, Discreteness of area and volume in quantum gravity, Nucl. Phys. B \textbf{442}, 593--619 (1995).

\bibitem{ref:lqg2}
A. Ashtekar and J. Lewandowski, Quantum theory of geometry: I. Area operators, Class. Quantum Grav. \textbf{14}, A55 (1997).

\bibitem{ref:dsr1}
G. Amelino-Camelia, Relativity in spacetimes with short-distance structure governed by an observer-independent (Planckian) length scale, Int. J. Mod. Phys. D \textbf{11}, 35--59 (2002).

\bibitem{ref:dsr2}
J. Magueijo and L. Smolin, Lorentz invariance with an invariant energy scale, Phys. Rev. Lett. \textbf{88}, 190403 (2002).

\bibitem{ref:Sabin}
S. Hossenfelder, {\em Classical and Quantum Gravity: Theory, Analysis and Applications}, Chap. 5, Edited by V. R. Frignanni, Nova Publishers (2011).

\bibitem{ref:Maggiro}
M. Maggiore, A Generalized Uncertainty Principle in Quantum Gravity, Phys. Lett. B \textbf{304},  65--69 (1993).

\bibitem{ref:Kempf1}
A. Kempf, Uncertainty relation in quantum mechanics with quantum group symmetry, J. Math. Phys. \textbf{35}, 4483 (1994).

\bibitem{ref:Kempf2}
A. Kempf, G. Mangano and R.B. Mann, Hilbert space representation of the minimal length uncertainty relation, Phys. Rev. D \textbf{52}, 1108 (1995).

\bibitem{ref:Ali}
A. F. Ali, S. Das and E. C. Vagenas, Discreteness of space from the generalized uncertainty principle, Phys. Lett. B \textbf{678}, 497--499 (2009).

\bibitem{ref:Gomes}
M. Gomes and V. G. Kupriyanov, Position-dependent noncommutativity in quantum mechanics, Phys. Rev. D \textbf{79}, 125011 (2009).

\bibitem{ref:Bagchi}
B. Bagchi and A. Fring, Minimal length in quantum mechanics and non-Hermitian Hamiltonian systems, Phys. Lett. A \textbf{373}, 4307–-4310 (2009).

\bibitem{ref:Quesne}
C. Quesne and V. M. Tkachuk, Composite system in deformed space with minimal length, Phys. Rev. A \textbf{81}, 012106 (2010).

\bibitem{ref:Pedram}
P. Pedram, K. Nozari and S.H. Taheri, The effects of minimal length and maximal momentum on the transition rate of ultra cold neutrons in gravitational field, J. High Energy Phys. \textbf{2011}, 1 (2011).

\bibitem{ref:Amelino}
G. Amelino-Camelia, M. Arzano, Y. Ling and G. Mandanici, Black-hole thermodynamics with modified dispersion relations and generalized uncertainty principles, Class. Quantum Grav. \textbf{23}, 2585 (2006).

\bibitem{ref:Majhi}
B. R. Majhi and E. C. Vagenas, Modified dispersion relation, photon's velocity, and Unruh effect, Phys. Lett. B \textbf{725}, 477--480 (2013).

\bibitem{ref:Ashtekar}
A. Ashtekar, S. Fairhurst and J. L. Willis, Quantum gravity, shadow states and quantum mechanics, Class. Quantum Grav. \textbf{20}, 1031 (2003).


\bibitem{ref:Corichi1}
A. Corichi, T. Vukasinac and J. A. Zapata, Polymer quantum mechanics and its continuum limit, Phys. Rev. D \textbf{76}, 044016 (2007).

\bibitem{ref:Kunstatter}
G. Kunstatter, J. Louko and J. Ziprick, Polymer quantization, singularity resolution, and the $1/r^2$ potential, Phys. Rev. A \textbf{79}, 032104 (2009).

\bibitem{ref:Laddha}
A. Laddha and M. Varadarajan, Polymer quantization of the free scalar field and its classical limit, Class. Quantum Grav. \textbf{27}, 175010 (2010).

\bibitem{ref:Hossain}
G. M. Hossain, V. Husain and S. S. Seahra, Nonsingular inflationary universe from polymer matter, Phys. Rev. D \textbf{81}, 024005 (2010).

\bibitem{ref:Prieto}
J. Fernando Barbero G, J. Prieto and E. J. S. Villase{\~n}or, Band structure in the polymer quantization of the harmonic oscillator, Class. Quantum Grav. \textbf{30}, 165011 (2013).
\bibitem{ref:Majumder}
B. Majumder and S. Sen, Do the modified uncertainty principle and polymer quantization predict same physics? Phys. Lett. B \textbf{717}, 291--294 (2012).


\bibitem{ref:GUP-Polymer1}
M. A. Gorji, K. Nozari and B. Vakili, Polymeric quantization and black hole thermodynamics, Phys. Lett. B \textbf{735}, 62--68 (2014).

\bibitem{ref:GUP-Polymer2}
M. A. Gorji, K. Nozari and B. Vakili, Polymer quantization versus the Snyder noncommutative space, Class. Quantum Grav. \textbf{32}, 155007 (2015).

\bibitem{ref:Das}
A. F. Ali, S. Das, and E. C. Vagenas, Proposal for testing quantum gravity in the lab, Phys. Rev. D \textbf{84}, 044013 (2011).

\bibitem{ref:Nozari}
K. Nozari and A. Etemadi, Minimal length, maximal momentum, and Hilbert space representation of quantum mechanics, Phys. Rev. D \textbf{85}, 104029 (2012).

\bibitem{ref:Igor}
I. Pikovski et al., Probing Planck-scale physics with quantum optics, Nature Phys. \textbf{8}, 393--397 (2012).

\bibitem{ref:Amelino2}
G. Amelino-Camelia, J. Ellis, N. E. Mavromatos, D. V. Nanopoulos and S. Sarkar, Tests of quantum gravity from observations of [gamma]-ray bursts, Nature \textbf{395}, 525 (1998).

\bibitem{ref:Vanner}
M. R. Vanner, J. Hofer, G. D. Cole and M. Aspelmeyer, Cooling-by-measurement and mechanical state tomography via pulsed optomechanics, Nature Commun. \textbf{4}, 2295 (2013).

\bibitem{ref:Amart2016}
F. Armata et al., Quantum and Classical Phases in Optomechanics, Phys. Rev. A \textbf{93}, 063862 (2016).

\bibitem{ref:Bosso}
P. Bosso, S. Das, I. Pikovski and M. R. Vanner, Amplified transduction of Planck-scale effects for quantum optical experiments,
Phys. Rev. A \textbf{96}, 023849 (2017).

\bibitem{ref:Kippenberg}
T.~J. Kippenberg and K.~J. Vahala, Cavity optomechanics: back-action at the mesoscale,
\newblock {Science} \textbf{321}, 1172--1176 (2008).

\bibitem{ref:Connell}
A.~D. O’Connell~et al., Quantum ground state and single-phonon control of a mechanical resonator,
\newblock {Nature} \textbf{464}, 697--703 (2010).

\bibitem{ref:Teufel}
J.~D. Teufel~et al., Sideband Cooling Micromechanical Motion to the Quantum Ground State,
\newblock {Nature} \textbf{475}, 359--363 (2011).

\bibitem{ref:Grossardt}
A.~Gro{\ss}ardt, J.~Bateman, H.~Ulbricht and A.~Bassi, Optomechanical test of the Schr{\"o}dinger-Newton equation,
\newblock {Phys. Rev. D} \textbf{93}, 096003 (2016).

\bibitem{ref:Gan}
C.~C. Gan, C.~M. Savage and S.~Z. Scully, Optomechanical tests of a Schr{\"o}dinger-Newton equation for gravitational quantum mechanics,
\newblock {Phys. Rev. D} \textbf{93}, 124049 (2016).


\bibitem{Dey_NPB}
S.~Dey et al., Probing noncommutative theories with quantum optical experiments,
\newblock {Nucl. Phys. B} \textbf{924}, 578--587 (2017).

\bibitem{ref:Law1995}
C. K. Law, Interaction between a moving mirror and radiation pressure: A Hamiltonian formulation, Phys. Rev. A \textbf{51} 2537 (1995).

\bibitem{ref:Vanner2010}
M. R. Vanner et al., Pulsed quantum optomechanics, Proc. Nat. Acad. Sci. USA \textbf{108}, 16182--16187 (2011).


\bibitem{ref:Corbitt2007}
T. Corbitt et al., Optical dilution and feedback cooling of a gram-scale oscillator to 6.9 mK, Phys. Rev. Lett. \textbf{99}, 160801 (2007).

\bibitem{ref:Harris2008}
J. D. Thompson, et al., The sculpting of Jupiter's gossamer rings by its shadow, Nature \textbf{453}, 72 (2008).

\bibitem{ref:Verlot2008}
P. Verlot et al., Scheme to probe optomechanical correlations between two optical beams down to the quantum level, Phys. Rev. Lett. \textbf{102}, 103601 (2008).

\bibitem{ref:Groblacher2009}
S. Gr\"{o}blacher, K. Hammerer, M. R. Vanner, M. Aspelmeyer, Observation of strong coupling between a micromechanical resonator and an optical cavity field, Nature \textbf{460}, 724--727 (2009).

\bibitem{Kleckner}
D.~Kleckner~et al., Optomechanical trampoline resonators,
\newblock {Opt. Express} \textbf{19}, 19708--19716 (2011).

\bibitem{ref:Martin2014}
F. Marin ~et al., Investigation on Planck scale physics by the AURIGA gravitational bar detector, New J. Phys. \textbf{16}, 085012 (2014).

\bibitem{ref:Bawj2015}
M. Bawj ~ et al., Probing deformed commutators with macroscopic harmonic oscillators, Nature Commun. \textbf{6}, 7503 (2015).

\bibitem{ref:Kumar2017}
S. P. Kumar and M. B. Plenio, Experimentally feasible quantum optical tests of Planck-scale physics, arXiv:1708.05659.
\end{thebibliography}

\end{document}